\documentclass[aps,prb,showpacs,groupedaddress,superscriptaddress,onecolumn]{revtex4}

\usepackage{graphicx,amsmath}
\usepackage{color}
\begin{document}
\baselineskip 17pt

\title{Disorder Effects in Topological StatesㄩBrief Review of the Recent Developments}

\author{Binglan Wu}
\affiliation{College of Physics, Optoelectronics and Energy, Soochow University, Suzhou 215006, China}

\author{Juntao Song}
\affiliation{Department of Physics, Hebei Normal University, Hebei 050024, China}

\author{Jiaojiao Zhou}
\affiliation{College of Physics, Optoelectronics and Energy, Soochow University, Suzhou 215006, China}

\author{Hua Jiang}
\email[]{jianghuaphy@suda.edu.cn}
\affiliation{College of Physics, Optoelectronics and Energy, Soochow University, Suzhou 215006, China}

\date{\today}

\begin{abstract}
Disorder inevitably exists in  realistic samples, manifesting itself in various  exotic properties for the topological states.
In this paper, we summarize and briefly review work completed over the last few years, including our own, regarding recent developments
in several topics about disorder effects in topological states. For weak disorder, the robustness of topological states is
demonstrated, especially for  both quantum spin Hall states with $Z_2=1$ and size induced nontrivial topological insulators with $Z_2=0$.
For moderate disorder, by increasing the randomness of both the impurity  distribution and the impurity induced potential, the topological insulator states can be created from  normal metallic or insulating  states. These phenomena and their mechanisms are summarized.  For strong disorder, the disorder causes a metal-insulator transition. Due to their topological nature,  the  phase diagrams are much richer in topological state systems.  Finally, the trends in these areas of disorder research are discussed.
\end{abstract}


\maketitle

\section{Introduction}
Topological states, including  gapped topological insulators and gapless topological semimetals, have become a focus of the condensed matter research \cite{Prange,Kane2010,Qi2011,Shen,Fang2015,Qiao2016B,Ryu2016,Weng2016}. In condensed matter systems, the first well-known  insulating topological state is quantum Hall effect (QHE) under a strong magnetic field \cite{Klitzing1980}. Subsequently, the quantum anomalous Hall effect (QAHE), a topological state similar as the QHE but without magnetic field, was proposed in 1988 \cite{Haldane}  and  observed in 2013 \cite{Chang2013}.
In 2005, Kane et al made a great step in the topological states research. They  proposed the two-dimensional $Z_2$ topological insulator - quantum spin Hall effect (QSHE) in graphene, which extends the topological state  into the class of systems protected by time reversal symmetry \cite{Kane2005A,Kane2005B}. Soon afterwards, the concept of symmetry protected topological states is broaden into three-dimension \cite{Fu2007,Moore2007} and  other discrete symmetries \cite{Ryu2016,Fu2011}.  Besides the insulating systems, the topological states can also exist in  gapless systems \cite{Volovik,XWan2011}. More recently, topological semimetals, containing both Weyl semimetals and Dirac semimetals, were experimentally verified \cite{HDing, ZKLiu}, soon after they were predicted in the corresponding  materials \cite{HWeng,ZWang2012}. The topological states are different from normal metallic and insulating states because of the existence of nontrivial topological order, which originates from the global properties of all electrons below the Fermi energy \cite{TKNN} and can be characterized by various types of  topological invariant numbers \cite{Kane2010,TKNN,Kane2005B}.  Due to the  nontrivial topological order, corresponding gapless  states emerge on the surface, leading to numerous  exotic  properties in topological systems. These properties have been reviewed
in references \cite{Prange,Kane2010,Qi2011,Shen,Fang2015,Qiao2016B,Ryu2016,Weng2016}.

Experimentally, disorder is ubiquitous  because of the defects in manufacturing processes and usually plays a dominant role in the transport properties of the samples being studied. Due to their unique electric structures, the response of topological states to  disorder is fundamentally different from that in  normal metals and insulators\cite{Prange,Qi2011,Shen,Klitzing1980,TKNN,Molekamp2007,TZhang2009,Jiang2009A,Jiang2014A,Jiang2012,Chen2015A,YQLi2010,HZLu2011,Aji2012,NPong2015}.  Physically, the topological invariant number  can  take on only a few discrete values. Therefore, the topological order cannot be easily disrupted by weak perturbations.  That is to say,  quantized transport that is robust against weak disorder can be obtained in various of topological systems. Moreover, arising from their  topological nature, topological states also show unconventional properties under moderate and strong disorder. Thus,  studies of disorder effects can not only give a comprehensive understanding about topological states, but also  offer a route map for the application of topological states.  In this paper, based on several of our finished works in the last few years \cite{Jiang2009A,Jiang2014A,Jiang2012,Chen2015A}, we  briefly review three topics covered by  recent studies of disorder effects in topological states.  We note that there are other topics related to disorder, such as the weak anti-localization \cite{YQLi2010,HZLu2011}, the chiral anomaly effects \cite{Aji2012,NPong2015} etc, which are not reviewed here.

The rest of this paper is organized as follows. In Section 2, the robustness of topological states against weak disorder is reviewed.  In section 3, we present how various type of disorder create topological insulator states from normal insulating or metallic states. In section 4, the main results from studies of the  metal-insulator transition in  topological states system are summarized. Finally, a brief conclusion and an outlook for future disorder research are given  in Section 5.

\section{Robustness of topological states}
Searching for material states with low-power dissipation is one of the  greatest challenges in  modern fundamental physics and materials science. In a real experimental sample,  disorder  always exists to some degree. A carrier propagating inside the sample will inevitably collide with  disorder sites and be scattered.  However, whether such a collision can cause energy dissipation depends on the ability of the scattering to induce  backscattering processes -- carriers propagating in the forward channel are scattered into the backward channel. In a normal metal, due to the spatial overlap between the forward and backward channels, backscattering can occur even for weak disorder. Therefore, the resistance (conductance) in a normal metal  is not universal but in general increases (decreases) with the disorder strength.  Transport in such a system is dissipative. In  sharp contrast,  in  various kinds of topological systems, the  spatial separation between the forward and backward channels and the existence of a special discrete symmetry (e.g. time-reversal symmetry) forbid  backscattering \cite{CKXu2006}.  As a result,  carriers  can propagate without dissipation. Thus, the resistance (conductance)  in these  systems is quantized and is insensitive to  the presence of weak disorder.   Such exotic phenomenon, namely the robustness of topological states, makes their host materials ideal  platforms for testing fundamental physics principles and achieving low-power dissipation in electronic/spintronic devices.

That topological states are robust against  weak disorder  was first discovered in a milestone experiment on quantum Hall effects \cite{Klitzing1980}. Klitzing et al observed quantized Hall resistance and zero longitudinal resistance, independent of  the material details.  Subsequently, such phenomena  have been extensively studied  both within the context of quantum Hall effect and in various  other topological states\cite{Datta2006,Qiao2009,LShen2005,Pan2015,Li2014,Shi2009,YYZhang2014}. In the  following, as examples, we demonstrate  the robustness of the topological states  and the mechanisms of this behavior in  two recently discovered systems: (i) quantum spin Hall effects (QSHE) in HgTe/CdTe quantum wells \cite{Jiang2009A} and (ii)  new proposed $Z_2=0$ topological insulators with emerging robust helical surface states\cite{Jiang2014A}.
\begin{center}
\includegraphics[width=0.7\columnwidth,viewport=200 0 880 540, clip]{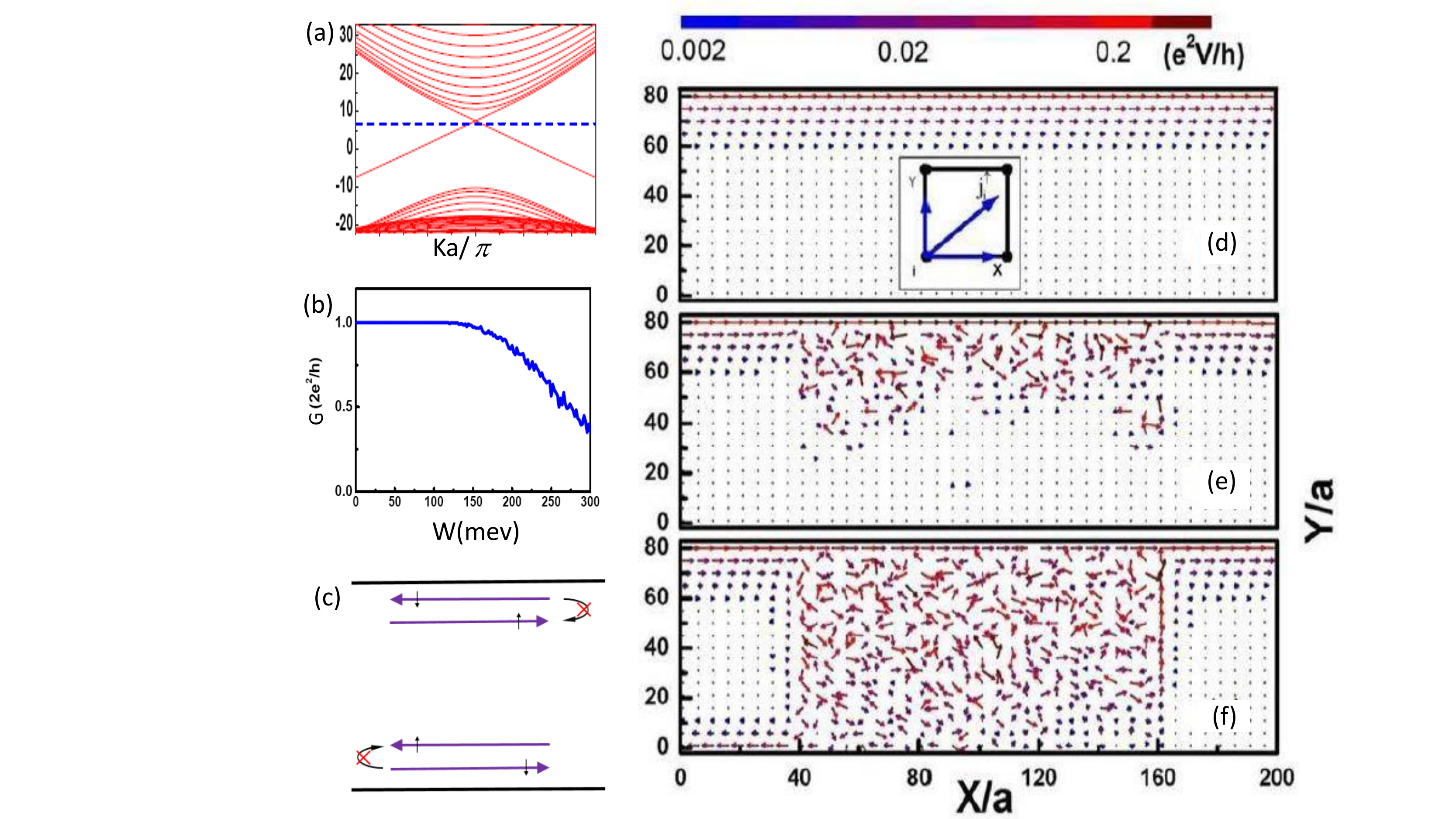}\\[5pt]  
\parbox[c]{15.0cm}{\footnotesize{\bf Fig.~1.} (color online) 
(a) Typical  energy spectrum for a HgTe/CdTe quantum well strip in the QSHE region.  (b) The conductance $G$ vs disorder strength $W$ in a two terminal device. The Fermi energy is located inside the inverted gap.  (c) Schematic of helical edge states propagation in the boundary of sample.  On a given edge, the carriers with opposite spin polarizations propagate in opposite directions.    (d) -(f) Configurations of the local current flow vector in a device with central region size $L_x= 200a, L_y=80a $ under disorder strength (d) $W=0$,(e) $W=110 {\rm meV}$, and (f) $W=220 {\rm meV} $.  The inset of  (d) is the schematic of local current flow vector. The direction and length of the arrows represent the local current direction and  magnitude. The order of magnitude of the local currents are displayed in the color bar. Adapted from Ref. \cite{Jiang2009A}}
\end{center}

For HgTe/CdTe quantum wells of moderate thickness, due to the strong spin-orbit coupling in this material, the  system can enter a new topological phases--the QSH phase \cite{SCZhang2006,Qi2006,Molekamp2007}. In Fig. 1(a),  the  energy spectrum for the  quasi one-dimensional strip is plotted.  There exists an inverted bulk energy gap with two degenerate energy bands (helical edge states) that cross inside the gap. When the Fermi energy is located inside the gap, e.g. $\varepsilon_F=7 {\rm meV}$ , the QSH phase is established.  To show how the transport properties of QSH phase are influenced by the disorder, a device with two terminals and a disordered central region is considered.  Fig. 1(b)shows the two terminal  conductance $G$ versus the disorder strength $W$ for the state shown in Fig. 1(a). For a range of disorder strength $W \in [0 {\rm meV}, 150 {\rm meV}]$ , the conductance G takes the quantized value $2 e^2/h$. Such an observation means that the QSHE is robust against weak disorder.  Preliminarily, the behavior of the conductance  $G$ can be understood on the basis of  spatial distribution of the helical edge states, as shown in Fig. 1(c). In the absence of disorder, the  spin up   channel on the top  boundary and the spin down channel on bottom boundary give rise to the $G=2 e^2 /h$. In the presence of nonmagnetic disorder,  time-reversal symmetry forbids the backscattering on a given edge \cite{CKXu2006}. Since the backscattering between the edge channels on the opposite boundaries decays exponentially due to the spatial separation, $G$ shows a  plateau with $G = 2 e^2/h$ without fluctuations.

Next,in order to  illustrate  clearly    the evolution of the conductance $G$ versus disorder,  the typical distributions of local currents in the device  are plotted in Fig. 1(d)-(f). Note that only the local currents for spin-up subsystem are  demonstrated.  The local currents for the spin-down subsystem can be directly obtained by applying the time-reversal symmetry. In a clean sample[Fig. 1(d)], the local currents are mainly located on the upper edge and their values decay exponentially toward  the bulk. The local currents spread into the bulk and  the edge channel is broadened when  disorder is introduced [Fig. 1(e)]. However, not until the disorder strength $W$ exceeds the critical value $W_c$, can the spreading local current  connect with the bottom edge channels, which have the opposite chirality. At this point, effective backscattering [Fig. 1(f)] can occur, leading to a decrease in the conductance $G$ between the two terminals. These pictures explain why $G$ is robust against  weak disorder and how it is destroyed in the strong disorder limit.

\begin{center}
\includegraphics[width=0.7\columnwidth,viewport=10 -10 820 540, clip]{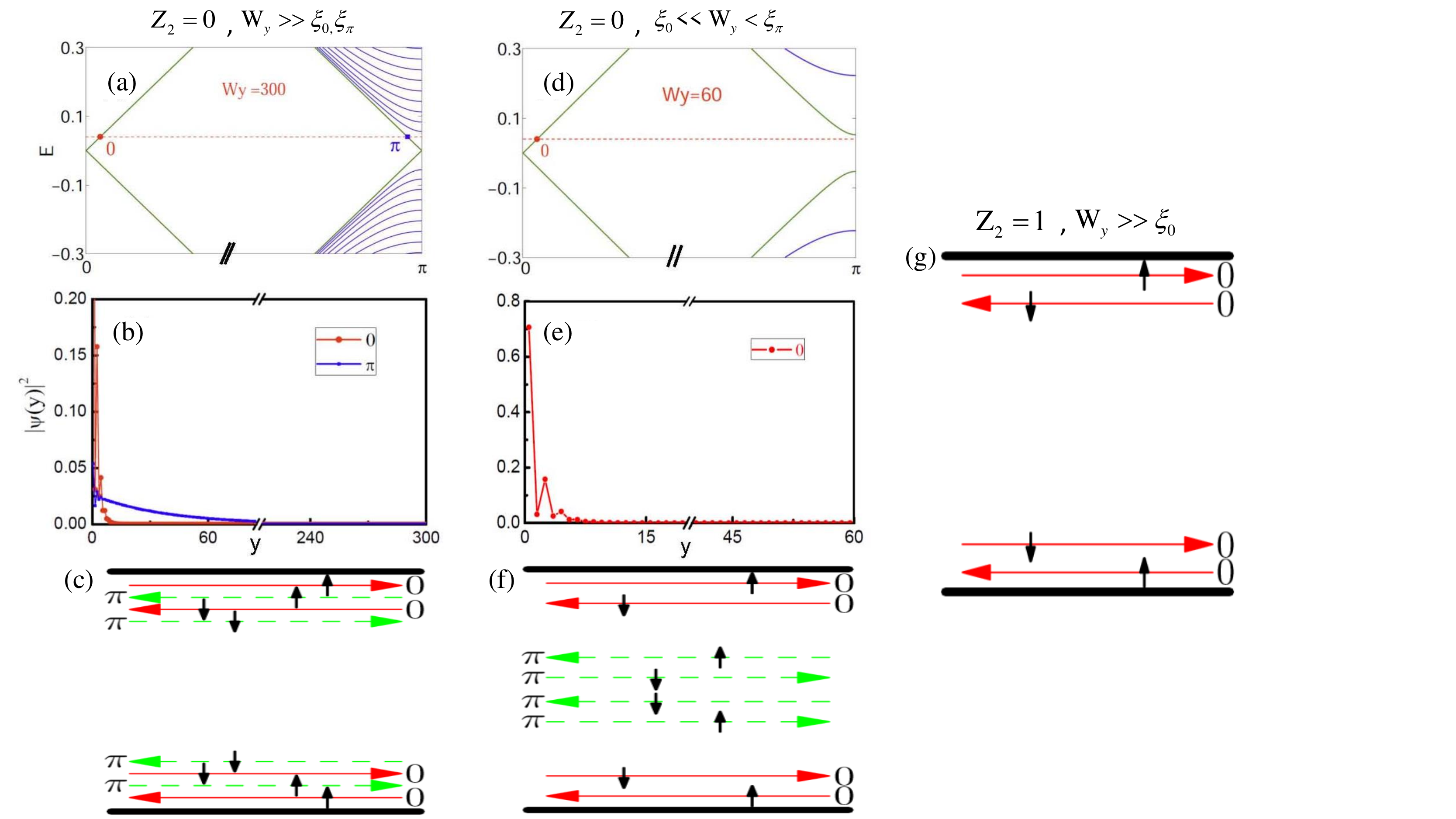}\\[5pt]  
\parbox[c]{15.0cm}{\footnotesize{\bf Fig.~2.} (color online)(a),(d) One dimensional energy bands for the  2D  anisotropic BHZ stripe with $Z_2=0$ and sample width ${\rm W_y}=300$ (a), 60 (d) described in the main text . (b),(e) show distribution of edge states in real space corresponding to (a) and (d) with fixed Fermi energy.(c) and (f) show schematic plots of helical edge modes corresponding to (a) and (d),  respectively. The vertical arrows represent the electron spin. In panel (f), the helical edge channels around $k_x= \pi$ are hybridized due to finite size confinement, leaving one pair of helical edge channels around $k_x=0$. Thus, these conducting channels resemble the helical edge channels in a $Z_2=1$ QSHE.  Panels (c),(f) and (g) are adapted from Ref. \cite{Jiang2014A}.  }
\end{center}

Topological insulators protected by time-reversal symmetry are characterized by the topological invariant number $Z_2$ \cite{Kane2005B}. There exist odd pairs of helical surface/edge states in  topological insulators with $Z_2=1$, which are robust and insensitive to material details and to external perturbations, as demonstrated in Fig. 1.  However,  most real materials with time reversal symmetry  have $Z_2=0$ topological order, and systems
with $Z_2 =1$ topological order are very rare. Indeed, the QSHE (two-dimensional topological insulator with $Z_2=1$) has been experimentally confirmed only in HgTe/CdTe \cite{Molekamp2007} and InAs/GaSb \cite{Du2011,Du2015} quantum wells. It is natural to ask, whether the $Z_2=1$ is a necessary requirement to realize topological surface/edge states.

According to the popular view, $Z_2=0$ topological systems have even pairs of helical surface/edge states, which are fragile in the presence of  disorder\cite{Shi2009,Kane2008}. However,  we find that for some  $Z_2=0$ systems,  robust transport can be engineered using a  combination of  finite size confinement and anisotropy \cite{Jiang2014A}. With the help of the anistropic Bernevig-Hughes-Zhang (BHZ) model \cite{SCZhang2006,Qi2006}, we demonstrate how  robustness against disorder arises in such topological states.

In an anisotropic, $Z_2 = 0$  BHZ stripe, for  certain parameters, there are two pairs of helical edge states around $k = 0$ and $k= \pi$ [Fig. 2(a)]. The $0$ helical edge states are located close to the edge, while the $\pi$ helical edge states are more delocalized toward the bulk [Fig. 2(b)]. When the stripe is sufficiently wide, [i.e. $W_y = 300$, which is much greater than the decay length of both the $0$ and $\pi$ helical edge channels], both $k = 0$ and $\pi$ helical edge bands are nearly gapless as expected. As a consequence, the Fermi energy crosses the $k = 0$ helical edge states and counter-propagating $\pi$ edge states, which also have substantial real-space overlap[Fig.2(c)]. The disorder can heavily couple these two states and cause strong backscattering between the $k = 0$ and $k= \pi$ edge channels, leading to complete localization. In transport experiments, such a  stripe then resembles  a normal insulator. In contrast, when the stripe becomes narrow [i.e. $W_y = 60$], the $0$ helical edge bands remain almost gapless, while the coupling of extended $\pi$ helical edge states opens a remarkable hybridization gap $\Delta$ [Fig. 2(d)]. When the Fermi energy is located inside the gap $\Delta$, only the $k=0$ helical edge states exist. Therefore, the $k=0$ helical edge states are well separated with the $k=\pi$ states.  In addition, since the decay length of 0 helical edge states, $\xi_0$, is much shorter than the sample width $W_y$, the $k=0$ helical edge states on the top edge are also well separated with that on the bottom edge [Fig. 2(e) and (f)].  From a transport point of view, the conducting edge channels in Fig.2(f) [$Z_2=0$] are similar to those in Fig. 2(g) and Fig. 1(c) [$Z_2=1$].  Conclusively, then despite having a topological number $Z_2=0$ , the  helical edge states in Fig. 2(f)  emerge as being  robust.

In order to quantitatively assess the robustness of these emerging $Z_2=0$ helical edge states, their transport properties under nonmagnetic   disorder in a two terminal device and a $\pi$-bar device [Fig. 3(a) and 3(e)] are simulated. The results are demonstrated in Fig. 3(c) and
3(e), respectively.  For comparison, the transport properties of these two devices incorporating (i) normal $Z_2=0$ states with two pairs of helical edge states [Fig. 3(b) and 3(f)] and (ii) $Z_2 =1$ QSHE [Fig. 3(d) and 3(h)]  are also obtained.  In the case of emergent helical edge states, the two terminal conductance $G_{12,12}$  shows a quantized plateau $2e^2/h$ in the clean limit. When nonmagnetic disorder is introduced, the $2 e^2/h$ plateau remains unchanged without fluctuation [Fig. 3(c)]. Meanwhile, the nonlocal conductance  $G_{14,23}$ in the $\pi$-bar device shows well quantized plateaus at $4 e^2/h$, irrespective of details of the terminals   and the strength of the strong disorder  [Fig. 3(g)]. Therefore, the transport properties of the emergent helical states are completely different from normal $Z_2=0$ states cases, where $G_{12,12}$ and $G_{14,23}$ are found that are found that are fragile against disorder. Instead, the $G_{12,12}$  and $G_{14,23}$ cases behave exactly like a $Z_2=1$ QSHE system, see Figs. 3(d) and 3(h), and the experimental results in \cite{Du2015}.   The two-terminal perfect $2 e^2/h$ plateau and the $\pi$-bar perfect $4e^2/h$  plateau  plausibly provide  a transport definition of robust helical edge states in these $Z_2=0$ systems. In a $Z_2=1$ QSHE, the robustness of the edge conduction is derived from its intrinsic topological character. On the other hand, the robust helical edge states in  the $Z_2=0$ model  arise from the fact that the backscattering  is detuned by the finite size confinement.

Besides the  anisotropic BHZ model, the $Z_2=0$ topological systems involving  helical surface states  with emergent robustness have been extended into several 2D and 3D models \cite{Jiang2014A,Guo2014,Fukui2015,Leeuw2015}. In addition, two  systems built on realistic materials:  (i) Dirac semi-metal with appropriate width and thickness confinement \cite{Wang2015} and  (ii) a bismuth (111) film with a thickness of between 20 and 70 nm \cite{Jiang2014A,Jin2014} are proposed to host robust helical surface states.  Recent relevant experiments have shown the clues of the existence of such  topological states in the latter proposed  material system \cite{Jin2014,Qian2016}. It is worth noting that the proposed $Z_2=0$ systems have additional exotic properties not present in $Z_2=1$ QSHE, which can be utilized to fabricate  new topological devices \cite{Wang2015}.

\begin{center}
\includegraphics[width=0.7\columnwidth,viewport=10 220 820 537, clip]{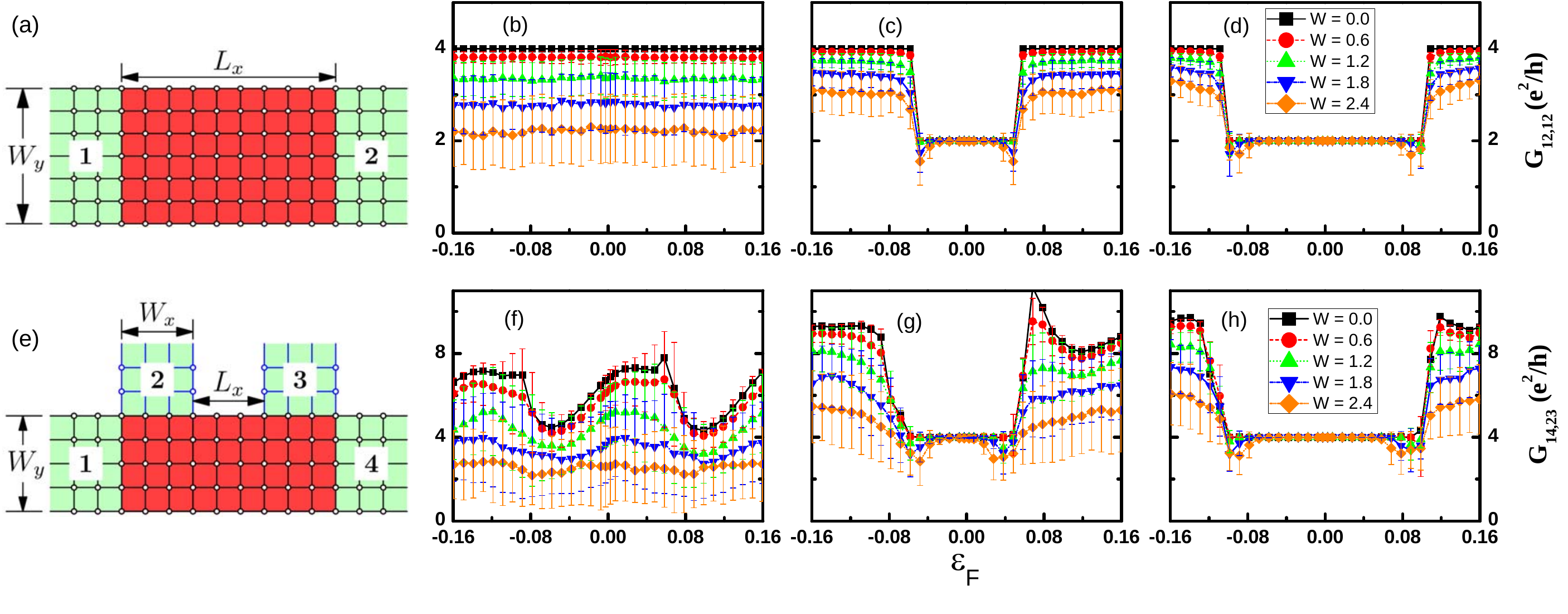}\\[5pt]  
\parbox[c]{15.0cm}{\footnotesize{\bf Fig.~3.} (color online)(a),(e) Illustration of two-terminal and $\pi$-bar devices . The Anderson disorder  exists only in the central (red) region. The size parameters are $L_x=120$, ${\rm W_x}=120$, ${\rm W_y}=60$. (b)每(d) Two-terminal conductance  $G_{12,12}$ of device (a); (f)每(h) nonlocal conductance  $G_{14,23}$ of device (e) vs the Fermi energy $\varepsilon_F$ for different disorder strengths $W$ for the cases of a normal $Z_2=0$ system (b) and (f);  $Z_2=0$ topological system with emergent robust helical edge states (c) and (g) and $Z_2=1$ QSHE (d) and (h).  $G_{14,23}$ means that the  current $I_{14}$  is injected from terminal 1 to  4 and  the voltage $V_{23}$ is measured between terminal 2 and 3 and $G_{14,23}=\frac{V_{23}}{I_{14}}$. The error bars denote the conductance fluctuation.  Adapted from ref. \cite{Jiang2014A}}
\end{center}

\section{Disorder induced Topological  insulators}
Besides destroying  topological states, in special systems, moderate disorder can also produce topological states.  The first well studied disorder induced topological state was the topological Anderson insulator (TAI). Specifically, in the clean limit,  the system is in an ordinary insulator or metal. With the introduction of  disorder, it enters into a topological insulator state with robust transport. This  TAI was  predicted to arise in HgTe/CdTe quantum wells by Shen's group \cite{Li2009} and our group \cite{Jiang2009A}.  This type of TAI  was then extended to many other systems \cite{Guo2009,Refael2015,Xing2011,DHXu2014,DHXu2016,ZYZhang2016,Borchmann2016,Shinsei2016,JTSong2014A,JTSong2014B,CPOrth2016,XRWang2016}.

Let us begin with a review of  how TAI arises in HgTe/CdTe quantum wells. In the clean limit, the system is described by the BHZ model\cite{SCZhang2006}
\begin{eqnarray}
\mathcal{H_0}(\vec{k})&=& \left(
                          \begin{array}{cc}
                            h_0(\vec{k}) & 0 \\
                            0 & h_0^{*}(-\vec{k}) \\
                          \end{array}
                        \right),  \nonumber\\
h_0(\vec{k}) & = & \varepsilon( \vec{k}) + d_\alpha (\vec{k}) \sigma_\alpha ,
\label{Equation1}
\end{eqnarray}
where $\sigma_\alpha$ [$\alpha=x,y,z$] are the Pauli matrices, and
\begin{eqnarray}
\varepsilon( \vec{k}) &=& C - D(k_x^2+k_y^2), \nonumber\\
\mathbf{d} (\vec{k}) &=& [Ak_x, Ak_y, M-B(k_x^2+k_y^2)].
\label{Equation2}
\end{eqnarray}
$A,B,C,D,M$ are material parameters. Specially, $M$ is the topological mass, which can be tuned continuously through the thickness of the HgTe layer.
If $M< 0$, the bulk band is inverted and the system  becomes to a QSHE [Fig. 4(b)]. In contrast, if $M >0$, the system becomes a normal insulator [Fig. 4(e)].

In Fig. 4(a) and 4(d), which is taken from Ref \cite{Li2009}ㄛ Li et al plot the two terminal conductance $G$ versus the disorder strength $W$ for different Fermi energies $\varepsilon_F$. When the Fermi energy $\varepsilon_F$ is in the valence band, then for both $M<0$ and $M > 0$, $G$ quickly decreases to  zero as the  disorder strength $W$ increases [green line in 4(a) and 4(d)]. In  sharp contrast, if $\varepsilon_F$ is located near the edge of the conduction band, for both $M<0$ and $M>0$ $G$ first decreases when Anderson disorder is introduced. Intriguingly, continuously increasing  the disorder strength $W$ does not lead to the localization of the system, but  makes $G$ increase again up to a quantized value. This value is maintained for a certain range without  fluctuations [blue line in 4(a) and 4(d)]. As described in  section II, the $2e^2/h$ plateau signals that the system has entered  a new topological state, referred above as the TAI. Further, the nature of the TAI is  clarified by the phase diagrams shown in Fig. 4(c) [$ M< 0$]  and in Fig. 4(f) [$M >0 $]. Significantly, for $M>0$, there are no topological helical edge states in the clean limit [Fig. 4(e)]. The quantized plateau cannot be attributed to the coexistence of the bulk and edge states, and the disorder causes the localization of the bulk states. It therefore seems that Anderson disorder creates the  helical edge state, and leads to  robust transport.

\begin{center}
\includegraphics[width=0.7\columnwidth,viewport=0 0 1660 1130, clip]{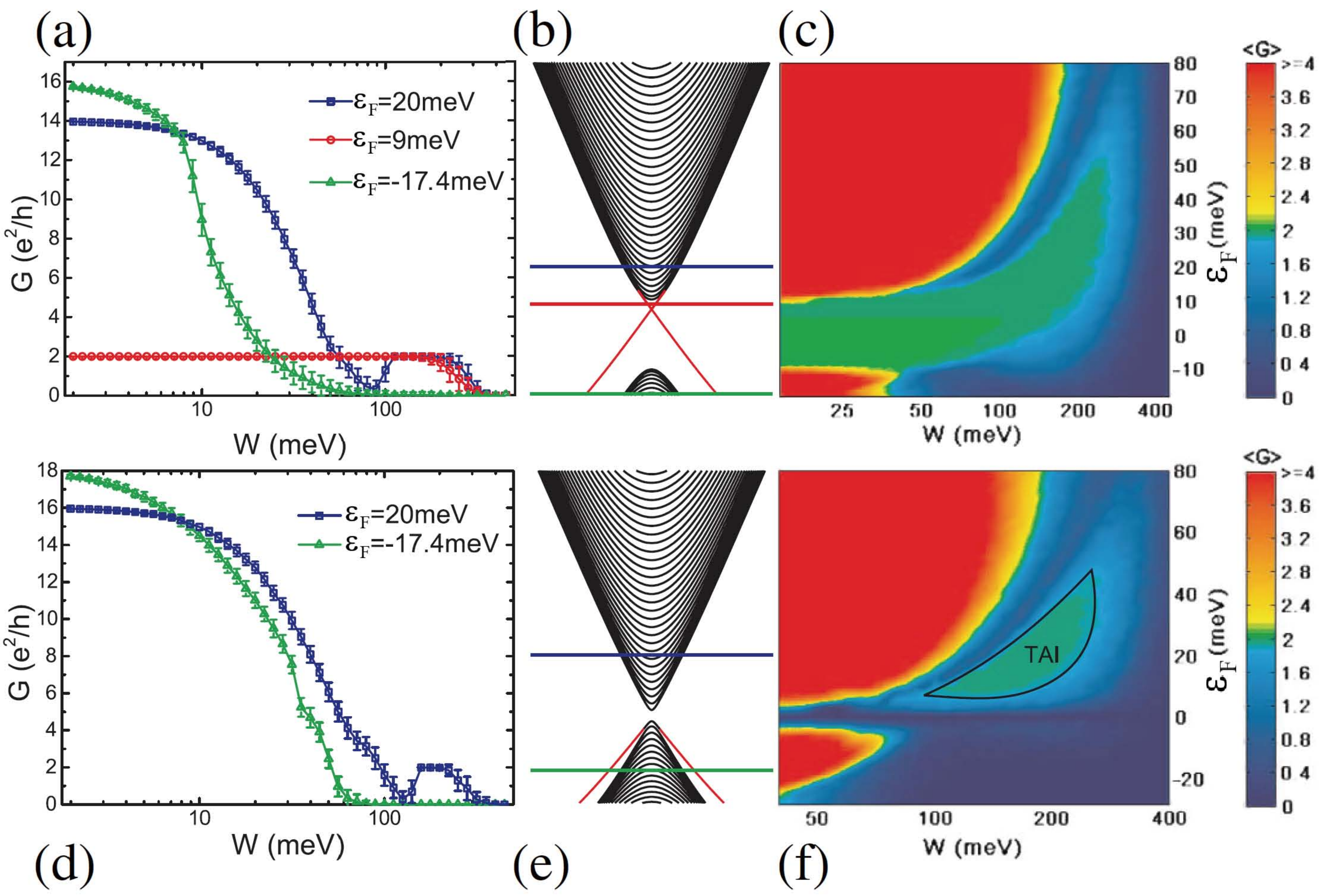}\\[5pt]  
\parbox[c]{15.0cm}{\footnotesize{\bf Fig.~4.} (color online) Conductance of disordered strips of HgTe/CdTe quantum wells.  (a)-(c) and  (d)-(f) show results for an inverted quantum well  with $M =-10 {\rm meV}$ and for a normal quantum well with $M=1 {\rm meV}$, respectively. (a) The conductance $G$ versus disorder strength $W$ at three values of Fermi energy. The error bars  represent the  conductance fluctuations. (b) One dimensional energy spectrum. The vertical scale (energy) is the same as in (c) and the horizontal lines correspond to the values of the Fermi energy considered in (a). (c) Phase diagram showing the conductance $G$ as a function of both disorder strength $W$ and Fermi energy $\varepsilon_F$. The panels (d), (e), and (f ) are the same as (a),(b), and (c), but for $M>0$. The TAI phase  is shown in green region. In all panels, the strip width is  $500 {\rm nm}$ and the length is $5000 {\rm nm}$ in (a) and (d), and $2000 {\rm nm}$ in (c) and (f ).  Adapted from Ref. \cite{Li2009}.}
\end{center}

To validate this assumption, we study the evolution  of the local current configurations under different disorder strengths [Fig. 5]\cite{Jiang2009A}. The parameters were set to be $M =2 {\rm meV}$ and $\varepsilon_F =18 {\rm meV}$ [Fig. (5a)].  The plot of $G$ vs $W$ in Fig. 5(b) can be subdivided into four regions (i) without disorder, (ii) before the anomalous plateau, (iii) on the anomalous plateau, and (iv) after the anomalous plateau.  Typical spatial distributions of the local currents are plotted in Fig. 5(h)-(k), respectively. The most interesting phenomena are to be found in  region (iii) [Fig. 5(i)]. Here, the local currents in the bulk decline to zero while the residual currents flow around the upper edge  with little scattering [only the spin up subsystem is considered]. In addition, throughout  region (iii), the bulk transport vanishes  and the edge transport shows the same behavior as in the traditional QSHE region [Fig. 1(d)-(f)]. The local currents shown in Fig. 5 provides strong evidence that  disorder leads to the helical edge states and hence to the TAI.

Based on the discoveries described in  references \cite{Li2009,Jiang2009A}, Groth et al. present an effective medium theory that explains the physical origin of the TAI\cite{Beenakker2009}.
Disorder can induce a self-energy $\Sigma$. In the self-consistent Born approximation (SCBA), $\Sigma$ is given by \cite{Bruus}:
\begin{eqnarray}
\Sigma = \frac{W^2}{12} (a/2\pi)^2 {\rm \int_{BZ}} d \vec{k} \Big{[}\sigma_\alpha \Big{(} \varepsilon_F^{+} -h_0(\vec{k})- \Sigma \Big{)}^{-1} \sigma_\alpha \Big{]}.
\label{Equation3}
\end{eqnarray}
Here $\sigma_\alpha$ denotes the type of disorder, and $a$ is the lattice constant.   $\Sigma$ can be decomposed into the form $\Sigma=\Sigma_0 \sigma_0 +\Sigma_x \sigma_x
+\Sigma_y \sigma_y+\Sigma_z \sigma_z$. Therefore, the self energy,  $\Sigma$,  makes a correction to the original Hamiltonian $h_0(\vec{k})$ and renormalizes the topological mass and
the Fermi energy:

\begin{eqnarray}
\overline{M} = M + {\rm Re}\Sigma_z , \ \ \  \overline{\varepsilon_F} = \varepsilon_F + {\rm Re} \Sigma_0.
\label{Equation3}
\end{eqnarray}

\begin{center}
\includegraphics[width=0.7\columnwidth,viewport=10 10 975 550, clip]{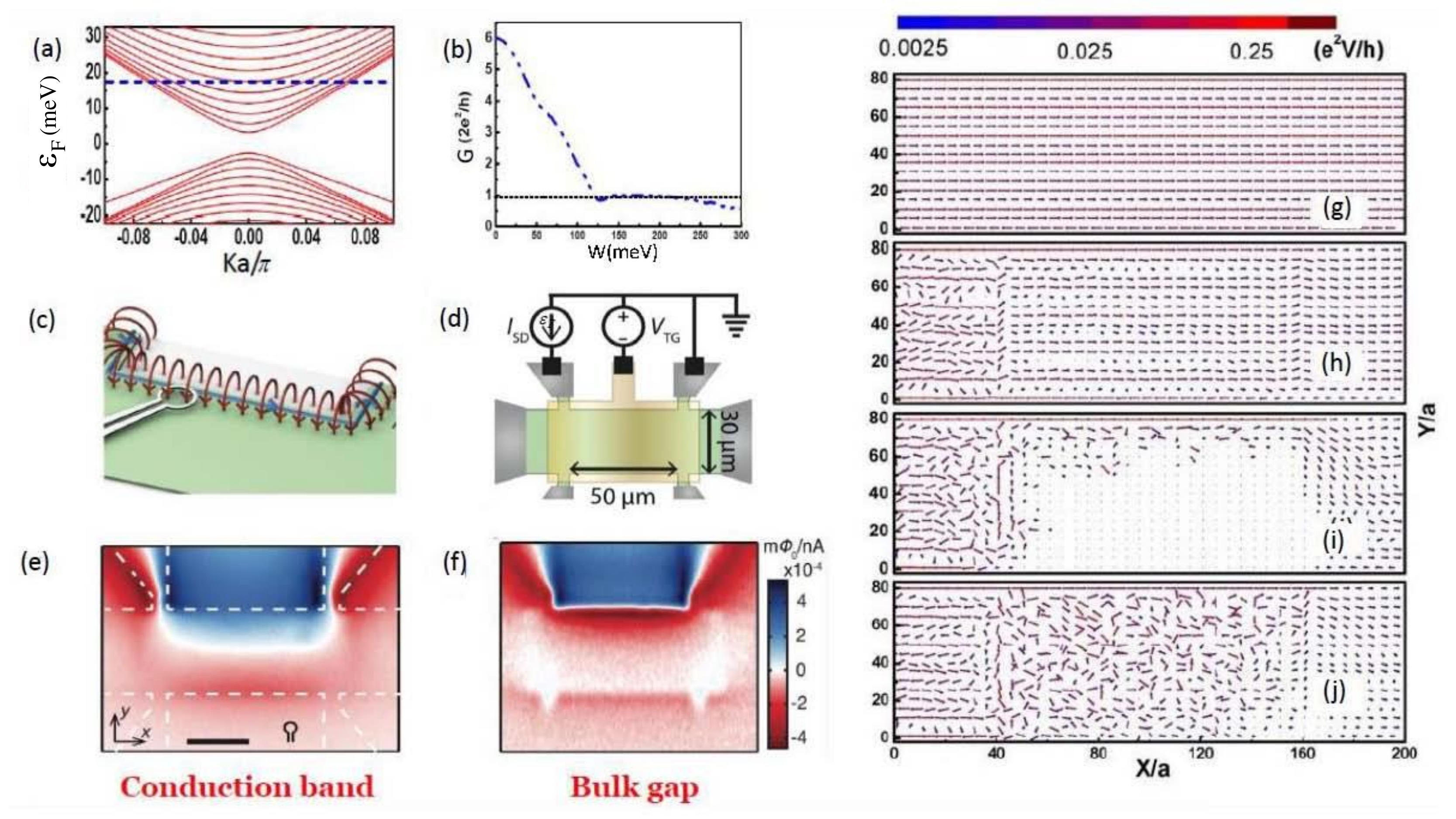}\\[5pt]  
\parbox[c]{15.0cm}{\footnotesize{\bf Fig.~5.} (color online ) (a) Typical one-dimensional energy spectrum for a normal HgTe/CdTe quantum well strip.  (b) Conductance $G$ vs disorder strength $W$.  (g)-(j) Configurations of the local current flow vector for the strip  with the same sample sizes as for Fig. 1, and with the  disorder strength set at (g) $W=0$, (h) $W=100 {\rm meV}$, (i) $W=150 {\rm meV}$, and (j) $W=250 {\rm meV}$. In (a),(b),(g)-(j),  The Fermi energy and topological mass are set to be $\varepsilon_F=18 {\rm meV}$  and  $M=2 {\rm meV}$. (c) Sketch of the configurations for the local current flow measurement.
The magnetic field (red) generated by the current (blue) is measured by detecting the flux through the SQUID's pickup loop.  (d) Schematic of a Hall bar. (e)(f) Typical measured distributions of the local current when Fermi energy is inside the conduction band (e) and the bulk gap (f), respectively.
(a)(b)(g)-(j) are reproduced from Ref. \cite{Jiang2009A}. (c)-(f) are adapted from Ref. \cite{Nowack}. }
\end{center}

Groth et al \cite{Beenakker2009} found that the  Anderson type of disorder [$\sigma_{0,z}$] can make a negative correction [${\rm Re}\Sigma_z <0$] to the topological mass $M$. Thus, it can lead a transition of $\overline{M}$ from positive to negative by increasing the disorder strength. Moreover, due to lacking the particle-hole symmetry of $h_0(\vec{k})$, the Anderson disorder also makes a negative correction to the Fermi energy $\varepsilon_F$.  When $\varepsilon_F$ is in the conduction band, for both $M>0$ and $M<0$, the Anderson disorder can  tune the renormalized Fermi energy $\overline{\varepsilon_F}$ into the inverted gap [-$|\overline{M}|$, $|\overline{M}|$]. The TAI phase is then established. In contrast, when $\varepsilon_F$ is in the valence band, $\overline{\varepsilon_F}$ is shifted away from the gap. $G$ quickly decays to zero and no quantized conductance appears.

If the effective medium theory is correct, one can deduce from Eq.(3) that the emergence of TAI is highly relevant to the type of disorder. Concretely, the $\sigma_z$ term exists in $\Big{(} \varepsilon_F^{+} -h_0(\vec{k})- \Sigma\Big{)}^{-1}$. While  Anderson disorder ($\sigma_{0,z}$) makes a negative correction  ($\sigma_0\sigma_z \sigma_0 = \sigma_z \sigma_z\sigma_z=\sigma_z$) to the topological mass $M$, the bond disorder ($\sigma_{x,y}$) results in a positive correction ($\sigma_x\sigma_z \sigma_x = \sigma_y \sigma_z\sigma_y=-\sigma_z$) to $M$.  To see this more clearly, in Fig. 6, we compare the conductance phase diagrams  for Anderson disorder  and bond disorder in the HgTe/CdTe quantum wells with $M =1~{\rm meV} $\cite{JTSong2012}.  It can be seen from Fig. 6(a), that at moderate Anderson disorder and with an  appropriate Fermi energy, a clear TAI phase [green region] is present, and such a TAI phase  should correspond to a negative renormalized topological mass $\overline{M}$ [blue region in Fig. 6(c)]. When  referring to the bond disorder case, for any disorder strength $W$ and Fermi energy $\varepsilon_F$, $\overline{M}$ is always  positive. Correspondingly, there is no indication of the TAI phenomenon in Fig. 6(b).  From  Fig. 6, one can conclude that normal Anderson disorder gives rise to the TAI phenomenon and bond disorder destroys it. The results not only verify the validity of the  effective medium theory, but also are important for the experimental realization of the TAI phase.

To data, the physical properties of the TAI phase have been  extensively studied~\cite{Prodan2011A,DWXu2012,YYZhang2012,YYZhang2013,Grischik2013,Grischik2015,Chen2012}, and several significant advances are worth noting. In Fig. 4(f) and Fig. 6(a), since it is surrounded by a  normal Anderson insulator phase, the TAI phase was initially considered to be a new topological phase. Later, using  a three dimensional phase diagram [disorder strength W, Fermi energy $\varepsilon_F$ and topological mass $M$], Prodan found that the  TAI pahse is not a distinct phase but part of the quantum spin-Hall phase, because these two phases can be adiabatically  connected by varying the topological mass \cite{Prodan2011A}.  Xu et al studied the phenomenon in a very large sample, and  a complete phase diagram of the TAI phase was obtained \cite{DWXu2012}. They found that the TAI phase region was much lager than what SCBA theory predicted. The reason is that the TAI can exist in a mobility gap rather than an inverted band gap \cite{DWXu2012,YYZhang2012}.

\begin{center}
\includegraphics[width=0.8\columnwidth,viewport=80 0 900 540, clip]{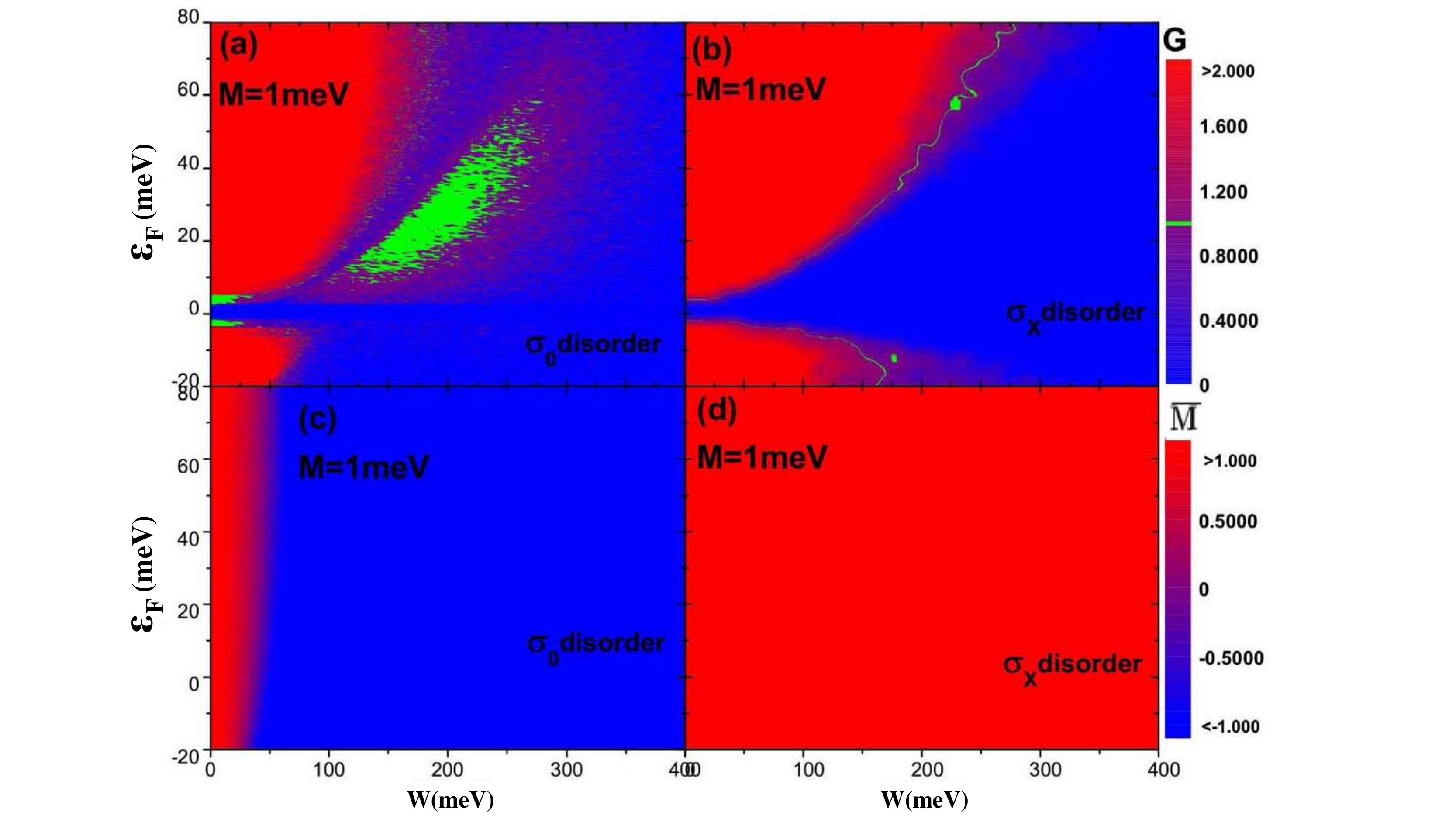}\\[5pt]  
\parbox[c]{15.0cm}{\footnotesize{\bf Fig.~6.} (color online) The conductance $G$ (a), (b) and renormalized topological mass $\overline{M}$ (c), (d) vs disorder strength $W$ and Fermi energy $\varepsilon_F$ for the HgTe/CdTe model with $M = 1 {\rm meV}$.  Here, $\sigma_0$ and $\sigma_x$ correspond to Anderson disorder and bond disorder, respectively. Adapted from Ref. \cite{JTSong2012}}
\end{center}

In what follows, we introduce another type of topological state induced by disorder.  Compared to the above TAI state, there are several differences. First, such a state is not triggered  by the disorder strength but the randomness of the adatom configuration. Second, such a state cannot be explained by the effective medium theory. Third and of the greatest importance, contrary to the common view, such disorder induced  topological states are much easier to be realized than the normal insulator state in the periodic case.

Because of the outstanding properties of its electronic structure, graphene is considered  as an ideal platform to host topological phases. For instance, recent  first-principles calculations proposed that certain nonmagnetic  adatoms (e.g., indium and thallium) could enhance the intrinsic  spin-orbit coupling to give rise to the QSHE \cite{Hu2011,Hu2012,Duan2013}; and  some magnetic adatoms (e.g., 3d and 5d transition metals) could induce  sizable Rashba spin-orbit coupling and magnetization to produce the quantized anomalous Hall effect (QAHE) \cite{Qiao2010,Qiao2011,HZhang2012}. Though great theoretical progress has been achieved, no experimental observation has been reported. There are two major reasons preventing the experimental exploration of these novel states. (i) All the calculations are based on the periodic adsorption condition, which is beyond  existing experimental capabilities. (ii) The formation of topological states is highly dependent on the adatom configurations. At certain coverage rates, the system is a trivial insulator.

Let us first explain why  the topological nontrivial state and trivial insulator state can exist in the periodic adsorption case with different coverage rates. As illustrated in Fig. 7(a) and (b), the tight-binding Hamiltonian of graphene with non-magnetic adatoms can be written as \cite{Qiao2012,Jiang2012}:
\begin{eqnarray}
H&=& H_0+ H_{\rm so} + H_{\rm U}                                       \nonumber \\
&=&- t \sum_{\langle i j \rangle,\alpha} c_{i\alpha}^{\dag}c_{j\alpha}+i \lambda_{\rm SO} \sum_{\langle\langle i j \rangle\rangle \in \mathcal{R} ,\alpha \beta} \nu_{ij}c_{i \alpha}^{\dag} s_{\alpha\beta}^{z} c_{j \beta}
+U \sum_{i \in \mathcal{R},\alpha} c_{i\alpha}^{\dag} c_{i\alpha}.
\end{eqnarray}

The presence of the adatom at $\mathcal{R}$  not only enhances the intrinsic SOC term ($H_{\rm so}$) but also generates the on site potential term ($H_{\rm U}$) with respect to the surrounding carbon atoms. $H_{\rm so}$ induces a topologically nontrivial gap $\Delta_{\rm so}$ at both $K$ and $K'$ with magnitude
\begin{eqnarray}
\Delta_{\rm so} & \propto & \sum_{\mathcal{R}} H_{\rm so}(\mathcal{R}) e^{i(K-K)\cdot R} = \sum_{\mathcal{R}} H_{\rm so}(\mathcal{R})  e^{i(K'-K')\cdot R} \nonumber\\
& = & \sum_{\mathcal{R}} H_{\rm so}(\mathcal{R})
\end{eqnarray}
and $H_{\rm U}$ induces the inter-valley scattering, which results in a topological trivial gap  $\Delta_{U}$ with magnitude

\begin{eqnarray}
\Delta_{ U} \propto \sum_{\mathcal{R}} H_{\rm U}(\mathcal{R}) e^{i(K-K')\cdot \mathcal{R}} .
\end{eqnarray}

When adatoms form a $\sqrt{3}n \times \sqrt{3} n $ or $3n \times 3n$ supercell, a finite $\Delta_U$ can be obtained since the factor $e^{i(K-K') \cdot \mathcal{R}} =1$ for  all $\mathcal{R}$. Because the on-site potential $U$ is much larger than the intrinsic SOC $\lambda_{\rm so}$, the topological trivial gap $\Delta_U$ exceeds the topological nontrivial gap $\Delta_{\rm so}$, resulting in a trivial insulator state. For other periodic adsorption cases, the inequivalent $e^{i(K-K') \cdot \mathcal{R}}$ for various $\mathcal{R}$ vanishes the inter-valley scattering [$\Delta_{U} \rightarrow 0$]. Since
$\Delta_{\rm so} > \Delta_{U}$, a quantum spin Hall state is realized \cite{Jiang2012}.

Interestingly, when the adatoms are  nonuniformly distributed in space, the factor $ e^{i(K-K') \cdot \mathcal{R}}$ becomes randomized. As a result, even when adatom is at the coverage rate $\frac{1}{3n^2}$ or $\frac{1}{9n^2}$, such randomization reduces $\Delta_U$  from  a large value in periodical case to nearly zero in the random adsorption case. We note the renormalization process for $\Delta_U$ cannot be described by the effective medium case. In  sharp contrast, the topologically nontrivial gap $\Delta_{\rm so}$ is caused by  zero momentum scattering. From Eq. (6),  the  randomization of the adatom distribution plays a negligible role in determining the magnitude of $\Delta_{\rm so}$.  Therefore, one can expect a quantum phase transition from trivial insulator state [$\Delta_{\rm so} < \Delta_{U}$]  to a topological insulator state [$\Delta_{\rm so} > \Delta_{U}$] with the introduction of the  spatial randomness of the adsorbates.

As an example, we study the possibility of realizing QSHE states in graphene with randomly distributed adsorbates.  Fig. 7(c) shows the results of a simulation of the transport
properties of the system with a  $11.1 \%$ coverage ratio [$3 \times 3$  supercell]. For the case of  periodically distributed adatoms [Fig. 7(a)], a zero
two terminal conductance $G=0$ is observed in the region where  $\varepsilon_F \in [0.117t, 0.124t]$, signaling a trivial insulator. However, when the adatoms become randomly distributed [Fig. 7(b)], a quantized plateau $G =2e^2/h$ with vanishing fluctuation emerges within the range  $\varepsilon_F \in [0.116t, 0.132t]$, indicating that the system turns into the  quantum spin Hall phase. Fig. 7(d) shows a simulation of the transport properties of the system with a $6.25 \%$ coverage ratio [$4 \times 4$  supercell].  The robust quantized plateau $G =2e^2/h$ for  both two cases shows that the topological phase predicted from  first-principles is not  affected by the randomization of the adatom configurations.  In Fig. 7(e), we show that randomization of magnetic adatoms can also turn graphene from a trivial insulator to a QAHE \cite{Qiao2012,Jiang2012}.

The above adatom adsorption studies suggest that in a realistic graphene sample, the topological state is the favored ground state. Therefore, this calculation  provides evidence of the high possibility of realizing topological phases in graphene  \cite{Jiang2012}.

\begin{center}
\includegraphics[width=0.7\columnwidth,viewport=20 53 960 485, clip]{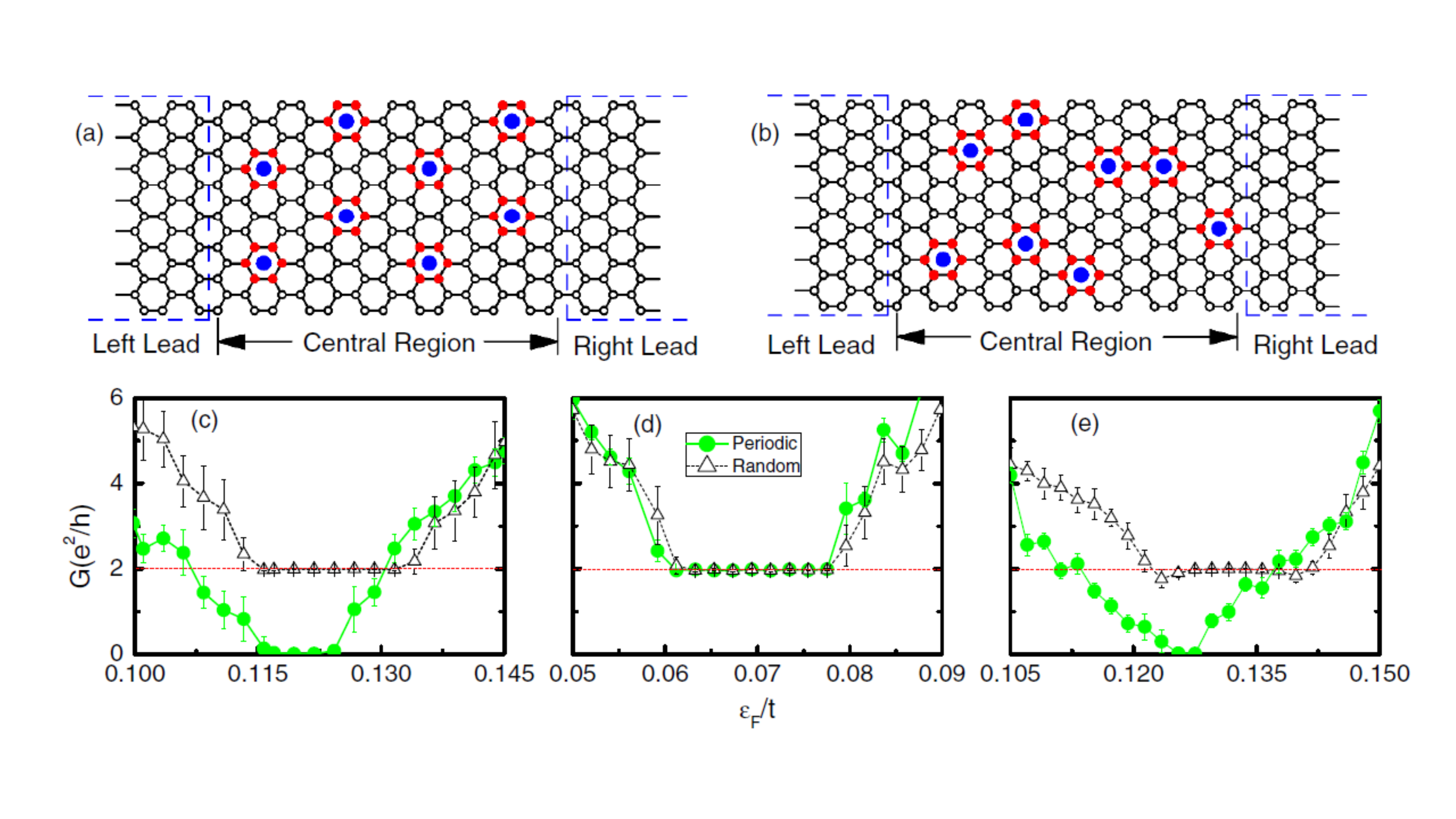}\\[5pt]  
\parbox[c]{15.0cm}{\footnotesize{\bf Fig.~7.} (color online)(a)(b) Schematic of a two-terminal device with periodically and randomly distributed adatoms, respectively. Both adatom coverages are $11.1\%$.(c)每(e) Comparison of conductances G between periodic and random adsorption as a function of Fermi level $\varepsilon_F$. (c)每(d) The site potential and intrinsic SOC are set to  $U = 0.36t$ and ${ \lambda_{\rm SO}} = 0.016t$. The adatom coverages are $11.1\%$ in panel (c) and $6.25\%$ in panel (d).  In (e)  the adatom coverage is $11.1\%$. Circle and triangle symbols represent the periodic and random adatoms. The error bars denote the conductance fluctuations. Adapted from Ref. \cite{Jiang2009A}}
\end{center}

\section{Metal-Insulator transition}
Since it was  first proposed by P. W. Anderson, the disorder induced metal-insulator transition has been a long lasting, interesting research issue in condensed matter physics \cite{Anderson1958}. According to  scaling theory, the metal-insulator transition in a material system depends on its universality ensemble [symplectic, unitary and orthogonal] and  dimension \cite{Anderson1979,Beenakker1997,Evers2008}. Initially, it was
believed that the  extended states (metal) could  exist only in  three-dimensional systems  and that all the bulk states in two and one-dimensional systems were  localized (insulator)  \cite{Anderson1979}. Later, two exceptions were found in two-dimensional systems. One was a the system with strong spin-orbital coupling [symplectic ensemble]. The other was a system without time reversal symmetry [unitary ensemble]. The topological states always share these two features and then become the focus of research on the  metal-insulator transition. In the past two decades, the metal-insulator transition in quantum Hall systems has been clearly understood~\cite{Huckestein,Chalker,Slevin}.  The extended state can only exist at the transition point between different plateaus. The critical exponents of the transition have  also been obtained.  In the last few years, there have been many studies  of  the metal-insulator transition in a QSHE\cite{Onoda2003,Nagaosa2007,Yamakaga2011,Yamakaga2013,Obuse2007}. Their results can be summarized as follows (iㄘ for a QSHE with Rashba spin-orbital coupling, the system belongs to a sympletetic ensemble. A metallic phase emerges between the QSH phase and  the normal insulator phase; (ii) for a QSHE without Rashba spin-orbital coupling, the system can be divided into two spin subsystems, both of which belong to unitary ensemble. A direct transition from a QSHE to a normal insulator is observed. Though the metal-insulator transition in two-dimensional topological states has been extensively studied,  the transition properties in three-dimensional topological states are still not quite clear.

More recently, the Weyl semimetal (WSM), a gapless topological state in three-dimension, has been theoretically predicted and experimentally confirmed \cite{XWan2011,HWeng,HDing}. Thus, it is timely to study the effects of disorder and metal-insulator transition both for their direct experimental relevance and for their fundamental value in advancing the understanding of the interplay between randomness and topological order. In reference \cite{Chen2015A}, we do so by using both numerical and analytical approaches. Different from the previous studies that only concentrate on single weyl node\cite{Brouwer2014,Brouwer2015,Pixley2015}, we focus on a more realistic tight-binding model with considering the interplay of opposite weyl nodes\cite{Nielsen,KYYang}.  By calculating the localization length \cite{MacKinnon1981} and the Hall conductivity \cite{JTSong2014,Prodan2011B}, we obtain the complete phase of the system under all the disorder strength. Because the WSM has novel gapless excitations, i.e. the Weyl nodes in the bulk and Fermi arcs on the surface, we find an unexpected rich phase diagram, as shown in Fig. 8(c).  With increasing disorder strength, the system undergoes multiple phase transitions. For example, one can obtain  (i) 3D quantum anomalous Hall state (QAH) -3D diffusive anomalous Hall metal (metal) - normal insulator (NI); (ii) WSM - QAH-metal-NI ; (iii) WSM-metal-NI; (iv) NI-WSM-metal-NI in Fig. 8(a),(b),(d),(e) respectively. Through the comprehensive study of the phase diagram, we  first address the important issue on the stability of the Weyl nodes and Fermi arcs against weak disorder.  The weak disorder has important effects when the Weyl nodes are close in momentum space. The WSM to QAH and the NI to WSM transitions are caused by the pair annihilation near the zone boundary and pair nucleation at the zone center.  Then, we find the the transition from WSM to metal by moderate disorder is unconventional. This transition takes place between two metallic  states and is only enabled by the topological character of the Weyl nodes.

\begin{center}
\includegraphics[width=0.7\columnwidth,viewport=0 13 980 460, clip]{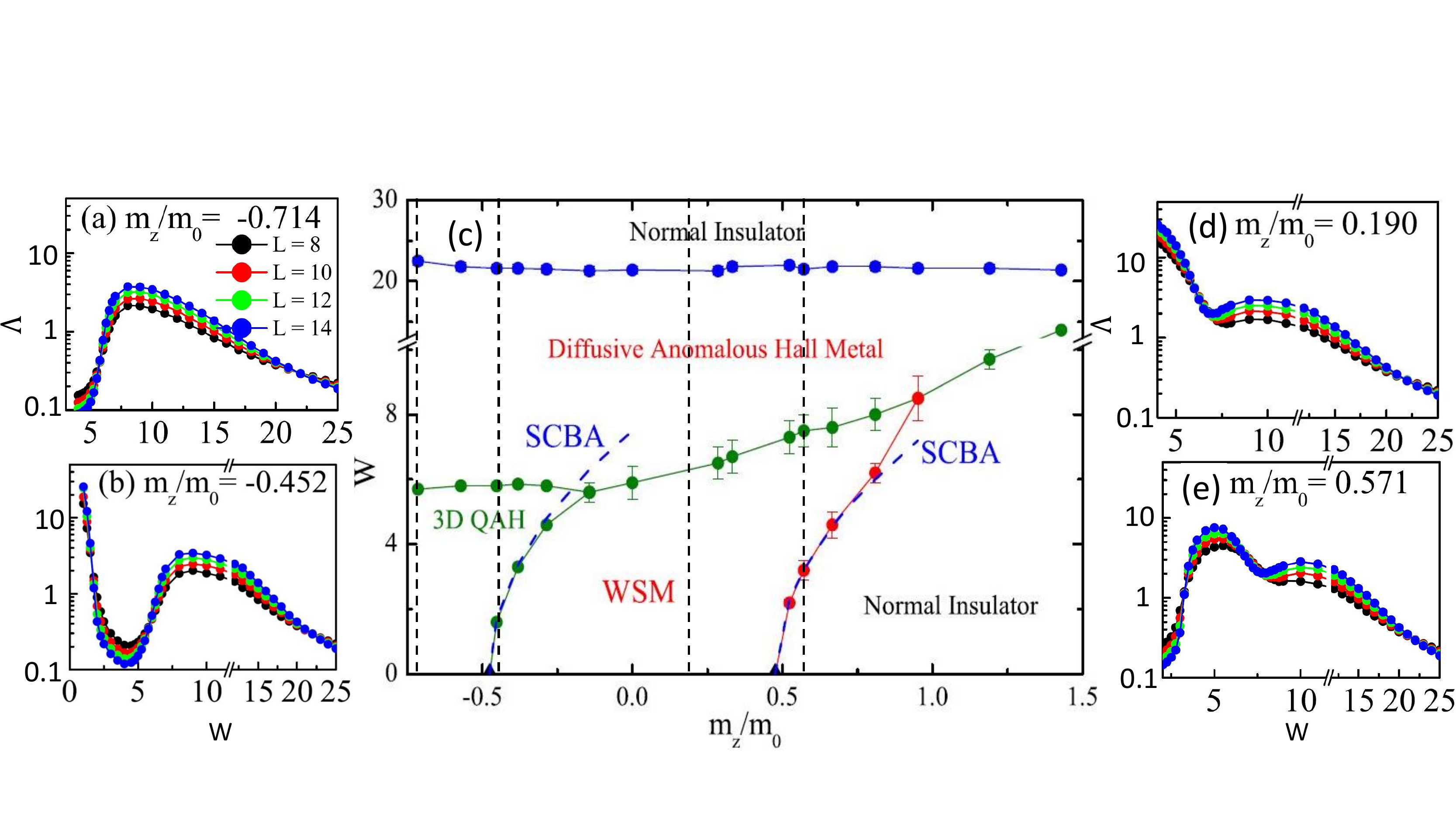}\\[5pt]  
\parbox[c]{15.0cm}{\footnotesize{\bf Fig.~8.} (color online) (a),(b),(d),(e) Normalized localization length $\Lambda= \lambda (L)/ L$ vs disorder
strength $W$ for different masses $m_z$ [lines in subplot (c)]. $\lambda (L)$ is the localization length of a long bar sample with cross section  $L \times L$.  An increase in $\Lambda$  with $L$ signals a metal phase, while a decrease with $L$ signals an insulator phase. When $\Lambda$ is independent of $L$, this signals a critical point of the phase transition. (c) Phase diagram in the $W-m_z$ plane. The symbols guided by the solid lines were obtained from the normalized localization length. The dashed blue  lines are the phase boundaries determined using the SCBA. In a finite layer sample, the 3D QAH, WSM and Diffusive anomalous Hall metal phases are distinguished by its Hall conductance being quantized and equivalent to to layer number, being quantized, and being non-quantized, respectively. Adapted from \cite{Chen2015A}}
\end{center}

It is worth noting that independent of our work, Liu et al also studied the disorder induced metal-insulator transition in 3D QAH layer systems and observed a rich  phase diagram \cite{Shindou2016}. Moreover, the unconventional WSM to metal transition has also generate broad interest, and the critical exponent of the transition has been  obtained \cite{Shindou2016,Hughes2016,JDSau2015}.

\section{Conclusions and outlook}

In this  paper,  based on reviewing our own disorder studies and other relevant works produced over  the last few years,  recent developments into the effects of disorder in  topological states are briefly summarized. We show that all of weak, moderate and strong disorder can lead to exotic phenomena in various type of topological states. How these phenomena originate from the topologically nontrivial nature of topological states is also demonstrated.  In spite of significant recent progress, there is  nevertheless still plenty of room for further research  into  the effects of disorder  in topological states. Before concluding the review, we discuss  the opportunities for  disorder studies in the future.

{\bf Experimental and material realization of disorder related topological phases}. Seven years after its first prediction \cite{Li2009,Jiang2009A},  the first signs of the experimental realization of the TAI phase have now been found. However, the host system is evanescently coupled waveguides \cite{Refael2015,Segev}.  Due to the recent great advance in controlling the disorder strength \cite{YQLi2015} and in the measurement of the local current \cite{Nowack} in  topological material, we expect the TAI phase can soon be confirmed in condensed matter systems. It is also to be noted that the topological phase in graphene has promising applications in information processing. We have demonstrated that both QSHE and QAHE states in graphene can easily be engineered through randomly adsorbing nonmagnetic/magnetic adatoms  \cite{Jiang2012}. We also expect the experimental observation of these topological states.

{\bf Understanding the fundamental phenomena caused by the disorder}. By now, our understanding of many  disorder related phenomena is still limited. Firstly, it is a common belief that 2D unitary system is scaled to insulator except at some isolated critical points \cite{Evers2008}. However, in reference \cite{Chen2015B} and reference \cite{Qiao2016}, though the considered models are totally different, we  find in both a novel  metallic phase region that  may emerge between QSHE  and normal insulator phases. The physical origins  for the metallic phase and the relationship between these two models should also be carefully addressed.  Secondly, analogue to disorder induced Anderson localization benefits the observation of quantized Hall plateaus in QHE,  the disorder also plays an important role in the experimental observation of the QSHE and QAHE \cite{Du2011,Du2015}. Du et al find that dilute silicon impurity doping in InAs/GaSb quantum wells can greatly suppress residual bulk conductance and produce a perfect QSHE \cite{Du2015}. From the point of view of theory, despite the attempt to study the phenomena\cite{DHXu2014}, the mechanism for the effects of impurity doping on the transport properties of the system is still not clear. Significantly, there exist  great puzzles regarding disorder effects in thin films of chromium-doped ${\rm (Bi,Sb)_2Te_3}$, which is  key to understand why the QAHE can only be observed at extremely low temperatures \cite{Yu2010,Chang2013}. Thirdly, type-II Weyl semimetals, which harbor unconventional Weyl nodes,  have  been proposed recently \cite{Dai2015}.  It would be interesting to study the stability of these Weyl nodes under strong disorder. Finally, dephasing effects also exist in  realistic samples. A natural topic is  how the transport properties of topological states are affected when both disorder and dephasing effects are considered\cite{Liu2014}.

{\bf Acknowledgments:}
We are grateful to  X. C. Xie, Q. Niu, Z. Q. Wang, Q.-F. Sun, J. R. Shi H. W. Liu, Z. H. Qiao,  J. Feng,
L. Wang, Z. B. Wang, D. W. Xu and C. Z. Chen for  collaboration and for their important contributions reviewed in this paper.
H. J. also appreciates the discussions with  J. Liu, J. H. Gao,  Y. G. Yao and S. Q. Shen. This Project supported by the National Natural Science Foundation of China (Grant No.11374219, No.  11474085 and No. 11534001)

\end{document}